\documentclass[useAMS,usenatbib,usedcolumn]{mn2e}
\usepackage{graphics,graphicx}     
\usepackage{amsmath}
\usepackage{amssymb}
\usepackage{natbib} 
\usepackage{ifpdf}

\title[MW companion discovered in first-year DES data]{Digging deeper into the Southern skies: a compact Milky-Way companion discovered in first-year Dark Energy Survey data}
\author[E. Luque et al.]{E. Luque,$^{1,2}$\thanks{\textbf{E-mail}: elmer.luque@ufrgs.br}  A. Queiroz,$^{1,2}$ B. Santiago,$^{1,2}$ A. Pieres,$^{1,2}$ E. Balbinot,$^{6,2}$  
\newauthor 
K.~Bechtol,$^{4}$ A. Drlica-Wagner,$^{5}$ A. Fausti Neto,$^{2}$ L. N. da Costa,$^{2,3}$  
\newauthor M. A. G. Maia,$^{2,3}$ B.~Yanny,$^{5}$ T.~Abbott,$^{7}$ S.~Allam,$^{5}$ A.~Benoit-L{\'e}vy,$^{8}$ 
\newauthor E.~Bertin,$^{9,10}$ D.~Brooks,$^{8}$ E.~Buckley-Geer,$^{5}$ D.~L.~Burke,$^{11,12}$ 
\newauthor A.~Carnero~Rosell,$^{2,3}$ M.~Carrasco~Kind,$^{13,14}$ J.~Carretero,$^{15,16}$ C.~E.~Cunha,$^{11}$ 
\newauthor S.~Desai,$^{17,18}$ H.~T.~Diehl,$^{5}$ J.~P.~Dietrich,$^{17,18}$ T.~F.~Eifler,$^{19,20}$ D.~A.~Finley,$^{5}$ 
\newauthor  B.~Flaugher,$^{5}$ P.~Fosalba,$^{15}$ J.~Frieman,$^{5,21}$ D.~W.~Gerdes,$^{22}$ D.~Gruen,$^{23,24}$ 
\newauthor G.~Gutierrez,$^{5}$ K.~Honscheid,$^{25,26}$ D.~J.~James,$^{7}$ K.~Kuehn,$^{27}$ N.~Kuropatkin,$^{5}$ 
\newauthor O.~Lahav,$^{8}$ T.~S.~Li,$^{28}$ M.~March,$^{19}$ J.~L.~Marshall,$^{28}$ P.~Martini,$^{25,29}$  
\newauthor R.~Miquel,$^{30,16}$ E.~Neilsen,$^{5}$ R.~C.~Nichol,$^{31}$ B.~Nord,$^{5}$ R.~Ogando,$^{2,3}$  
\newauthor A.~A.~Plazas,$^{20}$ A.~K.~Romer,$^{32}$ A.~Roodman,$^{11,12}$ E.~Sanchez,$^{33}$ V.~Scarpine,$^{5}$ 
\newauthor M.~Schubnell,$^{22}$ I.~Sevilla-Noarbe,$^{33,13}$ R.~C.~Smith,$^{7}$ M.~Soares-Santos,$^{5}$ 
\newauthor F.~Sobreira,$^{5,2}$ E.~Suchyta,$^{25,26}$ M.~E.~C.~Swanson,$^{14}$ G.~Tarle,$^{22}$ J.~Thaler,$^{34}$ 
\newauthor D.~Tucker,$^{5}$ A.~R.~Walker$^{7}$ and Y.~Zhang$^{22}$
\vspace*{.2cm}\\
Affiliations are listed at the end of the paper
}
\date{Released \today}
\pagerange{\pageref{firstpage}--\pageref{lastpage}} \pubyear{2014}

\def\LaTeX{L\kern-.36em\raise.3ex\hbox{a}\kern-.15em
    T\kern-.1667em\lower.7ex\hbox{E}\kern-.125emX}

\begin{document}
\newcommand{\mdash}{--}
\newcommand{\dash}{--}
\newcommand{\amp}{\&}
\newcommand{\araa}{ARA\&A}   
\newcommand{\afz}{Afz}       
\newcommand{\aj}{AJ}         
\newcommand{\azh}{AZh}       
\newcommand{\aap}{A\&A}      
\newcommand{\astap}{A\&A}      
\newcommand{\aaps}{A\&AS}     
\newcommand{\aapr}{A\&AR}     
\newcommand{\apj}{ApJ}       
\newcommand{\apjl}{ApJL}      
\newcommand{\apjs}{ApJS}     
\newcommand{\ao}{ApOpt}      
\newcommand{\cjaa}{Chinese J. Astron. Astrophys.}      
\newcommand{\apss}{Ap\&SS}   
\newcommand{\baas}{BAAS}     
\newcommand{\jaa}{JA\&A}     
\newcommand{\jrasc}{J. R. Astron. Soc. Can}   
\newcommand{\memras}{Mem. R. Astron. Soc.}  
\newcommand{\mensai}{Mem. Soc. Astron. Ital.} 
\newcommand{\mnras}{MNRAS}   
\newcommand{\nat}{Nat}       
\newcommand{\pasj}{PASJ}     
\newcommand{\pasp}{PASP}     
\newcommand{\paspc}{PASPC}   
\newcommand{\qjras}{QJRAS}   
\newcommand{\Sci}{Sci}       
\newcommand{\skytel}{Sky Telesc.} 
\newcommand{\sovast}{SvA}      
\newcommand{\ssr}{Space Sci. Rev.}   
\newcommand{\iaucirc}{IAU Circ.}  
\newcommand{\jaap}{JA\&A}
\newcommand{\pasa}{PASA}
\newcommand{\rasav}{Ric.~astr.~Specola astr.~Vatic.}
\newcommand{\sca}{Scient.~Am.}   
\newcommand{\stel}{Sky Telesc.}  
\newcommand{\spsrev}{Space Sci.~Rev.} 
\newcommand{\phfl}{Phys. Fluids}
\newcommand{\phrev}{Phys. Rev.}
\newcommand{\rprph}{Rep. Prog. Phys.}
\newcommand{\rmph}{Rev. Mod. Phys.}
\newcommand{\jplph}{J. Plasma Phys.}
\newcommand{\jmph}{J. Math. Phys.}
\newcommand{\jgeores}{J. Geophys. Res.}
\newcommand{\solphys}{Sol. Phys.}

\def\sun{\ensuremath{\odot}}
\def\farcs{\hbox{$.\!\!^{\prime\prime}$}}
\def\la{\mathrel{\mathchoice {\vcenter{\offinterlineskip\halign{\hfil
$\displaystyle##$\hfil\cr<\cr\sim\cr}}}
{\vcenter{\offinterlineskip\halign{\hfil$\textstyle##$\hfil\cr
<\cr\sim\cr}}}
{\vcenter{\offinterlineskip\halign{\hfil$\scriptstyle##$\hfil\cr
<\cr\sim\cr}}}
{\vcenter{\offinterlineskip\halign{\hfil$\scriptscriptstyle##$\hfil\cr
<\cr\sim\cr}}}}}
\def\ga{\mathrel{\mathchoice {\vcenter{\offinterlineskip\halign{\hfil
$\displaystyle##$\hfil\cr>\cr\sim\cr}}}
{\vcenter{\offinterlineskip\halign{\hfil$\textstyle##$\hfil\cr
>\cr\sim\cr}}}
{\vcenter{\offinterlineskip\halign{\hfil$\scriptstyle##$\hfil\cr
>\cr\sim\cr}}}
{\vcenter{\offinterlineskip\halign{\hfil$\scriptscriptstyle##$\hfil\cr
>\cr\sim\cr}}}}}
\def\deg{\hbox{$^\circ$}}
\def\arcmin{\hbox{$^\prime$}}
\def\arcsec{\hbox{$^{\prime\prime}$}}
\def\utw{\smash{\rlap{\lower5pt\hbox{$\sim$}}}}
\def\udtw{\smash{\rlap{\lower6pt\hbox{$\approx$}}}}
\def\fd{\hbox{$.\!\!^{\rm d}$}}
\def\fh{\hbox{$.\!\!^{\rm h}$}}
\def\fm{\hbox{$.\!\!^{\rm m}$}}
\def\fs{\hbox{$.\!\!^{\rm s}$}}
\def\fdg{\hbox{$.\!\!^\circ$}}
\def\farcm{\hbox{$.\mkern-4mu^\prime$}}
\def\farcs{\hbox{$.\!\!^{\prime\prime}$}}
\def\fp{\hbox{$.\!\!^{\scriptscriptstyle\rm p}$}}
\def\diameter{{\ifmmode\mathchoice
{\ooalign{\hfil\hbox{$\displaystyle/$}\hfil\crcr
{\hbox{$\displaystyle\mathchar"20D$}}}}
{\ooalign{\hfil\hbox{$\textstyle/$}\hfil\crcr
{\hbox{$\textstyle\mathchar"20D$}}}}
{\ooalign{\hfil\hbox{$\scriptstyle/$}\hfil\crcr
{\hbox{$\scriptstyle\mathchar"20D$}}}}
{\ooalign{\hfil\hbox{$\scriptscriptstyle/$}\hfil\crcr
{\hbox{$\scriptscriptstyle\mathchar"20D$}}}}
\else{\ooalign{\hfil/\hfil\crcr\mathhexbox20D}}%
\fi}}
\def\squareforqed{\hbox{\rlap{$\sqcap$}$\sqcup$}}
\def\sq{\ifmmode\squareforqed\else{\unskip\nobreak\hfil
\penalty50\hskip1em\null\nobreak\hfil\squareforqed
\parfillskip=0pt\finalhyphendemerits=0\endgraf}\fi}
\def\vec#1{\ensuremath{\mathchoice{\mbox{\boldmath$\displaystyle#1$}}           
{\mbox{\boldmath$\textstyle#1$}}                                                 {\mbox{\boldmath$\scriptstyle#1$}}                                               
{\mbox{\boldmath$\scriptscriptstyle#1$}}}}

\label{firstpage}

\maketitle

\begin{abstract}
Stellar substructure within the Milky Way's halo, including dwarf galaxies, star clusters, and streams, holds a wealth of information about the formation and evolution of our Galaxy. The detection and characterization of these structures require homogeneous deep imaging data over a large area of the sky. The Dark Energy Survey (DES) is a 5000 sq. degree survey in the southern hemisphere, which is rapidly reducing the existing north-south asymmetry in the census of MW satellites and other stellar substructure. We use the first-year DES data down to previously unprobed photometric depths to search for stellar systems in the Galactic halo,  therefore complementing the previous analysis of the same data carried out by our group earlier this year. Our search is based on a matched filter algorithm that produces stellar density maps consistent with stellar population models of various ages, metallicities, and distances over the survey area. The most conspicuous density peaks in these maps have been identified automatically and ranked according to their significance and recurrence for different input models. We report the discovery of one additional stellar system besides those previously found by several authors using the same first-year DES data. The object is compact, and consistent with being dominated by an old and metal-poor population. DES\,J0034-4902 is found at high significance and appears in the DES images as a compact concentration of faint blue point sources at $\simeq 87\,\mathrm{kpc}$. Its half-light radius of $r_h = 9.88\pm 4.31\, \mathrm{pc}$ and total luminosity of $M_V\sim -3.05_{-0.42}^{+0.69}$ are consistent with it being a low mass halo cluster. It is also found to have a very elongated shape. In addition, our deeper probe of DES 1st year data confirms the recently reported satellite galaxy candidate Horologium\,II as a significant stellar overdensity. We also infer its structural properties and compare them to those reported in the literature.
\end{abstract}


\section{Introduction}
The census of Milky Way (MW) satellites has grown rapidly over the past fifteen years. Several of these newly found objects are star systems with very low luminosities ($-3.0\lesssim M_V\lesssim 0$) and small half-light radii ($<10\,\mathrm{pc}$), being more consistent with star clusters \citep{Koposov2007, Belokurov2010, Fadely2011, Munoz2012, Balbinot2013}. 
These clusters are thought to be suffering stellar mass loss via dynamical processes such as tidal disruption or evaporation \citep{Koposov2007,Kim2015a}. 
The number of dwarf galaxies around the MW has also increased significantly, from the 12 classical dwarfs known until the late 1990s, up to a total of 27 which were known by early this year \citep{McConnachie2012}, thanks in large part to the Sloan Digital Sky Survey (SDSS). Several of these latter were shown to be very low luminosity systems with high $M/L$, thus representing some of the most dark matter rich objects.

At the larger luminosities typical of globular clusters (GCs), different cluster subpopulations classified by their position, kinematics and horizontal branch morphology have been known for several decades \citep{Zinn1985,Zinn1993}. The so-called young halo clusters may have originated in dwarf galaxies accreted by the MW \citep{Zinn1993,Lee1994}. Both types of objects seem to share a vast planar structure around the Galaxy, which also encloses several stellar and gaseous streams of clusters and dwarf galaxies \citep{Pawlowski2012,Pawlowski2015}. The accretion origin of part of the MW system of GCs is also supported by the fact that several of them are found to have positions and kinematics that relate them to the Sagitarius dwarf galaxy \citep{Law2010b}. 
On the other hand, at the much lower luminosities of the recent satellite discoveries, the very distinction between star clusters and dwarf satellites may become less clear, as attested by their respective loci in size and luminosity space. It is therefore important to pursue a complete census of faint stellar systems inhabiting the Galactic halo, and to characterize them in terms of structure, stellar populations and dark matter content. Extrapolations of the SDSS results over the entire sky and over the currently known luminosity function of MW dwarfs indicate that this census is still very incomplete \citep{Tollerud2008, Hargis2014}. 

A very recent boost to the number of known MW satellites has been brought by the 
Dark Energy Survey \citep[DES;][]{Abbott2005}. Using the first internal release of DES coadd data (Y1A1), \cite{Bechtol2015} reported on the discovery of eight new MW satellites over a solid angle of 1,800 square degrees in the southern equatorial hemisphere. Six of these systems have sizes and optical luminosities clearly consistent with the low-luminosity dwarfs previously detected in SDSS. The case for the other two objects is less clear. In a parallel effort, \cite{Koposov2015} reported nine new MW satellites using the same DES imaging data, including the same eight and one additional object. One of the objects in common between these two searches, Kim\,2, had in fact been previously found by \cite{Kim22015} using data from the Stromlo MW Satellite Survey. In addition to that, \cite{Kim2015b} have discovered yet another object using Y1A1 data, Horologium\,II. The Panoramic Survey Telescope and Rapid Response
System 1 (PAN-STARRS) has also been responsible for several recent discoveries of Milky-Way satellites \citep{Laevens2014,Laevens2015a,Laevens2015b}.

As described in \cite{Bechtol2015}, several complementary search strategies have been implemented within the DES Collaboration to search for stellar substructures. In \cite{Bechtol2015}, we used a conservative star selection to ensure high stellar purity and completeness as well as a uniform field density over the survey footprint. The present work extends the results presented in \citet{Bechtol2015} by including stars at fainter magnitudes and by considering a broader range of spatial extensions as well as ages and metallicites for the stellar populations composing new satellite systems. We also describe in detail the application of another search algorithm to the Y1A1 coadd data. Together, these analysis updates have enabled the discovery of a new candidate stellar cluster, DES\,J0034-4902,  which we call DES 1, and the confirmation of Horologium\,II as a physical stellar system. In \S \ref{sec:data} we describe the first-year DES data used. In \S \ref{sec:method} we describe the matched-filter algorithm applied to find the new systems. The new discovery is presented in \S \ref{sec:results}. In \S \ref{sec:HorII} we report on the detection and characterization of Horologium\,II.  Our final remarks are given in \S \ref{sec:conclusions}.

\section{DES Data}
\label{sec:data}
DES is a wide-field imaging program expected to cover about $\mathrm{5\,000\,deg^2}$ in the $grizY$ bands down to $\simeq 24.6^{th}$ magnitude (at $\mathrm{S/N}\simeq 10$ for galaxies  in $g$ band; \cite{Abbott2005}) in the southern equatorial hemisphere for a period of five years. It uses the Dark Energy Camera (DECam), a $\mathrm{3\,deg^2}$ ($2\fdg 2$ diameter) mosaic camera with $\mathrm{0\farcs 263}$ pixels on the prime focus of the CTIO Blanco 4\,m telescope \citep{Flaugher2015}. The DECam images are reduced by the DES Data Management (DESDM) team, which has developed a pipeline to process the data 
from basic single exposure instrumental corrections all the way to catalogue creation from calibrated coadded images.
Here we use the DES year one coadd catalogue data (Y1A1), taken from August 2013 to 
February 2014. For more details on Y1A1 and DESDM we refer to \citet{Sevilla2011}, \citet{Mohr2012}, and Gruendl et al., in prep. 
The stellar sample used in this work was drawn using the {\tt SExtractor}  parameters  ${\tt FLAGS}$, ${\tt SPREAD\_MODEL}$, and ${\tt PSF}$ magnitudes \citep{Bertin1996,Desai2012,Bouy2013}. These latter quantities are weighted average values taken from individual exposures of each source. We used a source quality criterion of ${\tt FLAGS} \leq 3$ over the $gri$ filters. 

As mentioned above, \cite{Kim2015b} have discovered one stellar object (Horologium\,II) in Y1A1 data that was not initially identified by \citet{Bechtol2015}, or by \cite{Koposov2015}. We believe that a primary reason for the non-detection of this object is that most of the probable member stars are fainter than $g\simeq 23\,\mathrm{mag}$, which is where \citet{Bechtol2015} set the faint-end threshold to search of stellar objects. This conservative threshold to select stars was set to ensure high stellar purity and completeness, as well as a uniform field density over the survey footprint. 

In this work, we adopt a selection in ${\tt SPREAD\_MODEL}$ intended to increase stellar completeness, specifically, $i$-band ${\tt |SPREAD\_MODEL|} < 0.003+ {\tt SPREADERR\_MODEL}$. A bright (faint) $g$ magnitude limit of ${\tt MAG\_PSF} = 17$ (${\tt MAG\_PSF} = 24$) was also applied. The faint limit is 1\,mag deeper than used by \cite{Bechtol2015}. In order to prevent point sources with extreme colours (including red dwarfs from the Galactic disk) from contaminating the sample, a colour cut at $-0.5 \leq g-r \leq 1.2$ was also used.

\section{Search Method}
\label{sec:method}
As discussed in \cite{Bechtol2015}, several independent search methods were used in the original analysis of Y1A1 data. In this section we describe in detail a different method, which was the one primarily used in this work.

\subsection{Matched Filter}
\label{sec:mf}

The Matched Filter (MF) technique has several applications for signal processing. In the context of astronomy, it has been used to detect low-density features and populations in imaging data \citep{Szabo2011,Rockosi2002}. We here use it to search for new star clusters and dwarf galaxies following on the work by \cite{Balbinot2011}.\par
The number of stars as a function of position on the sky ($\alpha$,$\delta$) and of colour ($c$) and magnitude ($m$) may be generally described as
\begin{equation}
n(\alpha,\delta,c,m) = n_{cl}(\alpha,\delta,c,m)+ n_{bg} (\alpha,\delta,c,m).\label{eq:1} 
\end{equation}
The first term on the right hand side corresponds to the contribution by the cluster ($cl$) we want to discover, whereas the second term represents the background ($bg$), which includes foreground halo stars and background unresolved galaxies. We then split these terms into a normalization term and a probability distribution function (PDF): 
\begin{equation}
n_{cl}(\alpha,\delta,c,m) = \zeta_{cl}(\alpha,\delta)f_{cl} (c,m),\label{eq:2} 
\end{equation}
where $\zeta_{cl}$ and $f_{cl}$ are the number normalization and PDF on the color-magnitude diagram (CMD) plane, respectively, for the stellar population to be found. The stellar population may be extended in space (as in a stream), but we explicitly  assume that its CMD is the same everywhere. As for the background stars, Galactic structure models show that both the number density and CMD vary as a function of position across the sky. So, we write
\begin{equation}
n_{bg}(\alpha,\delta,c,m) = \zeta_{bg}(\alpha,\delta)f_{bg} (\alpha,\delta,c,m).\label{eq:3} 
\end{equation}
With the definitions above, equation (\ref{eq:1}) then becomes
\begin{equation}
n(\alpha,\delta,c,m) = \zeta_{cl}(\alpha,\delta)f_{cl} (c,m) + \zeta_{bg}(\alpha,\delta)f_{bg} (\alpha,\delta,c,m). \label{eq:4}
\end{equation}
We bin stars into spatial pixels of area of $\mathrm{1\farcm 0 \times 1\farcm 0}$, indexed by $i$, and color-magnitude bins of $\mathrm{0.01\, mag \times 0.05\, mag}$, indexed by $j$. Details on the construction of the $f_{cl}$ and $f_{bg}$ PDFs are found in \S 3.2 and \S 3.3, respectively. With this notation,
\begin{equation}
n(i,j) = \zeta_{cl}(i)f_{cl} (j) + \zeta_{bg}(i)f_{bg} (i,j).\label{eq:5} 
\end{equation}
The left hand side is the expected number of stars in a given spatial pixel and CMD bin. If the actual number of stars observed in a catalog is $N(i,j)$, the variance between data and model is
\begin{equation}
s^2(i) = \sum\limits_{j}\frac{[N(i,j) - \zeta_{cl}(i) f_{cl}(j) - \zeta_{bg}(i) f_{bg}(i,j)]^2}{\zeta_{bg}(i)f_{bg}(i,j)}.\label{eq:6}
\end{equation}
The term in the denominator expresses the expected Poisson fluctuation in the star counts, which, for simplicity, we assume to be dominated by the background. Minimizing the variance and solving for $\zeta_{cl}(i)$, we have the number of observed stars that, according to the model given by equation (\ref{eq:4}), are consistent with the model. 
\begin{equation}
\zeta_{cl}(i) = \frac{\sum_{j}N(i,j)f_{cl}(j)/f_{bg}(i,j)}{\sum_{j}f_{cl}^2(j)/f_{bg}(i,j)} - \frac{\zeta_{bg}(i)}{\sum_{j}f_{cl}^2(j)/f_{bg}(i,j)}.\label{eq:7}
\end{equation}
The output of the filter application is a stellar density map of stars which are probable cluster members stars, i.e., $\zeta_{cl}(i)$. 
In practice, $f_{bg}(i,j)$ is generated from our target stellar catalog itself. We do that under the assumption that the contamination by any yet to be detected cluster, dwarf galaxy, or stellar stream, does not change the background PDF. As for the object PDF, we make use of simulated samples, as described in the next subsection.

\subsection{Model Grid}
\label{sec:grid}

\begin{table}
\caption{Parameter grid used to simulate SSPs for the search of star systems in DES Y1 footprint.}\centering
\scalebox{1.0}{
\begin{tabular}{lccc}\hline
Parameters & Lower limit & Upper limit & steps\\\hline
$\mathrm{\log(Age)}$ & $9.0$ & $10.2$ & $0.3$\\
$\mathrm{Distance\,(kpc)}$ &$10$ & $200$ & $10$ \\\hline
$\mathrm{Metallicity},\,Z$ & \multicolumn{3}{c}{$0.0002$,\, $0.001$\, and\, $0.007$}\\\hline
\end{tabular}}
\label{pargrid}
\end{table}

Since we do not know a priori what stellar populations we will find, we create a grid of simple stellar populations (SSPs) with the code {\tt GenCMD}\footnote{ https://github.com/balbinot/gencmd}. {\tt GenCMD} uses {\tt PARSEC} isochrones by \cite{Bressan2012} for different assumed distances and randomly selects stellar masses following a predefined initial mass function (IMF). Currently, we are adopting a \cite{Kroupa2001} IMF for that purpose. Given each stellar mass, we interpolate among the isochrone entries to draw absolute magnitudes in the desired filters. These are converted into {\it  measured} apparent magnitudes using the assumed model distance, magnitude uncertainties taken from Y1A1, and the reddening map of \cite{Schlegel1998}. Positions on the sky may also be simulated assuming different profile models. We simulate several SSPs at various ages, metallicities and distances covering a broader range of isochrone choices, including younger and higher metallicity stellar populations, than those adopted in \cite{Bechtol2015}. The parameter grid of these simulations is presented in Table \ref{pargrid}.

\subsection{Object Detection}
\label{Object finding}
We apply the MF method as presented in \S \ref{sec:mf} to the stellar catalog using each of the SSPs in the model grid described in \S \ref{sec:grid}. In practice, the sky is partitioned into $\sim\mathrm{10\degr \times 10\degr}$ cells to account for local variations in the background CMD, which is empirically derived from the stars in each individual cell. This procedure generates one density map for every sky cell and for every point in the model grid. \par
The maps are then convolved with Gaussian spatial kernels of different sizes \{$\sigma=0\farcm 0$ (no convolution) to $\sigma = 9\farcm 0$\} to highlight substructure on scales typical of star clusters and ultra-faint dwarf galaxies. In particular, smaller spatial kernels are suitable for the detection of more compact stellar systems. Our choice of spatial kernel sizes is such as to complement the ones adopted by \cite{Bechtol2015}.

As it is not practical to visually inspect all the resulting maps from the large number of combinations of sky cells, SSP models, and spatial convolution kernels, we use the {\tt SEXtractor} code to automatically search for density peaks. In fact, the convolution kernels are applied from within {\tt SEXtractor} itself as we run it on maps of different sky regions resulting from different SSP models.
Any object found by {\tt SEXtractor} in each map is recorded. Objects are then ranked according to the number of times they are detected by {\tt SEXtractor}. This is done separately for each sky cell and convolution kernel. 
The {\tt SEXtractor} parameters for the search were defined as those that maximized the recovery of simulated objects of different sizes and richness inserted into the DES stellar catalogue. 

\begin{figure}
\centering
\includegraphics[width=0.42\textwidth]{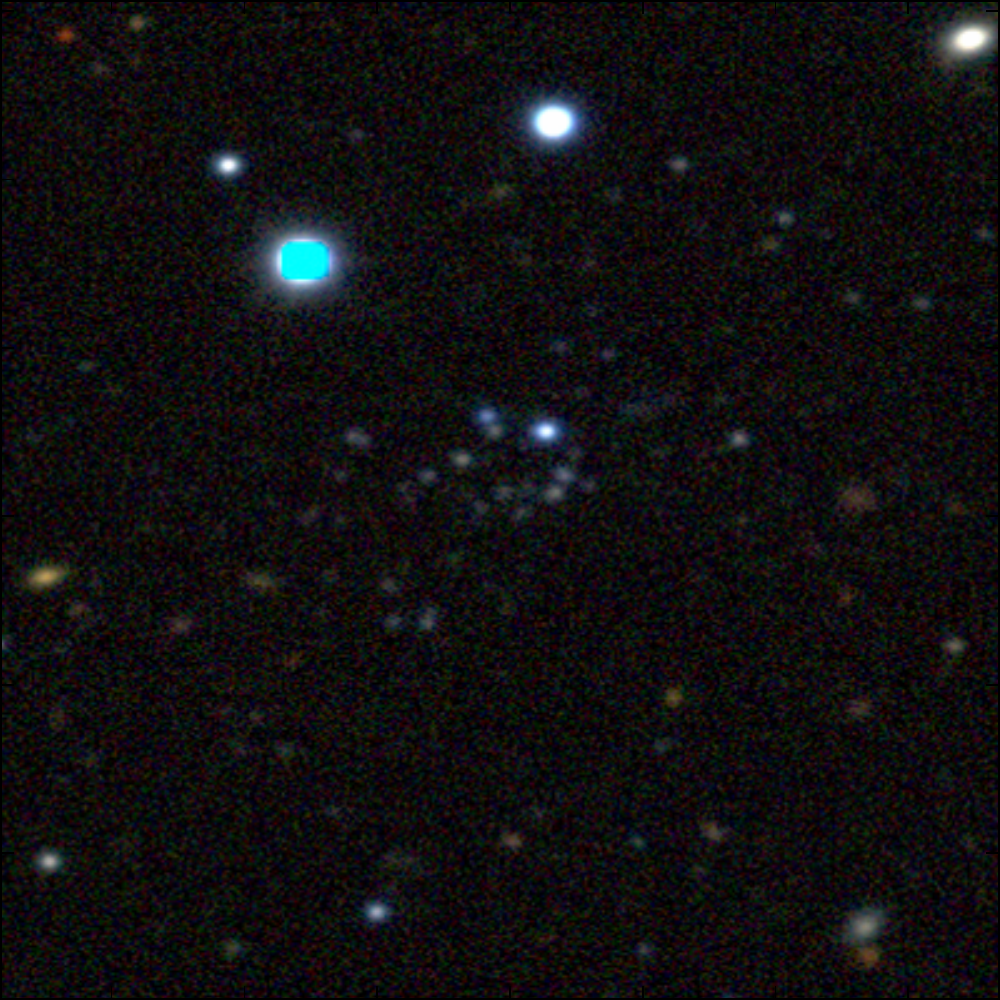}\vspace*{.3cm}
\caption{DES coadd image cutout of DES\,1 taken from the DES Science portal. The image is $1\farcm 78\times 1\farcm 78$ centered on DES\,1. The R,G,B channels correspond to the $i,\,r,\,g$ bands.}
\label{J0034tv}
\end{figure}

The first ten objects in each region of the sky and for each convolution kernel are visually inspected to identify the most likely candidates. We visually checked the stellar density map around them, the Poisson statistical significance above the background represented by their associated stellar peak, their number density profile and CMD. The density maps, significance and density profiles provide a basic assessment of the overdensities being found. The CMD helps us judge if this overdensity is consistent with a stellar population. All these diagnostic tools are shown in the next section for DES\,1.

We validated the detection method described above, which we call {\tt SparSEx}, by applying it to simulated SSPs superposed on real SDSS and DES data. We also tried to recover faint MW satellites previously discovered in SDSS data. In particular, we have chosen seventeen objects found in SDSS data (eleven dwarf galaxies and six star clusters) that are characterized as distant and ultra faint objects. {\tt SparSEx} detected all these stellar objects at the top of the object ranking lists.
 The rate of success for simulated stellar systems was the same. Finally, {\tt SparSEx} detected all eight satellites reported by \cite{Bechtol2015} and Horologium\,II reported by \cite{Kim2015b}. This later is discussed in more detail in \S \ref{sec:HorII}. The ninth object detected by \cite{Koposov2015} is in a region of Y1 data that is not included in the Y1A1 coadd that we searched.

\begin{figure*}
\begin{center}
\includegraphics[width=.95\textwidth]{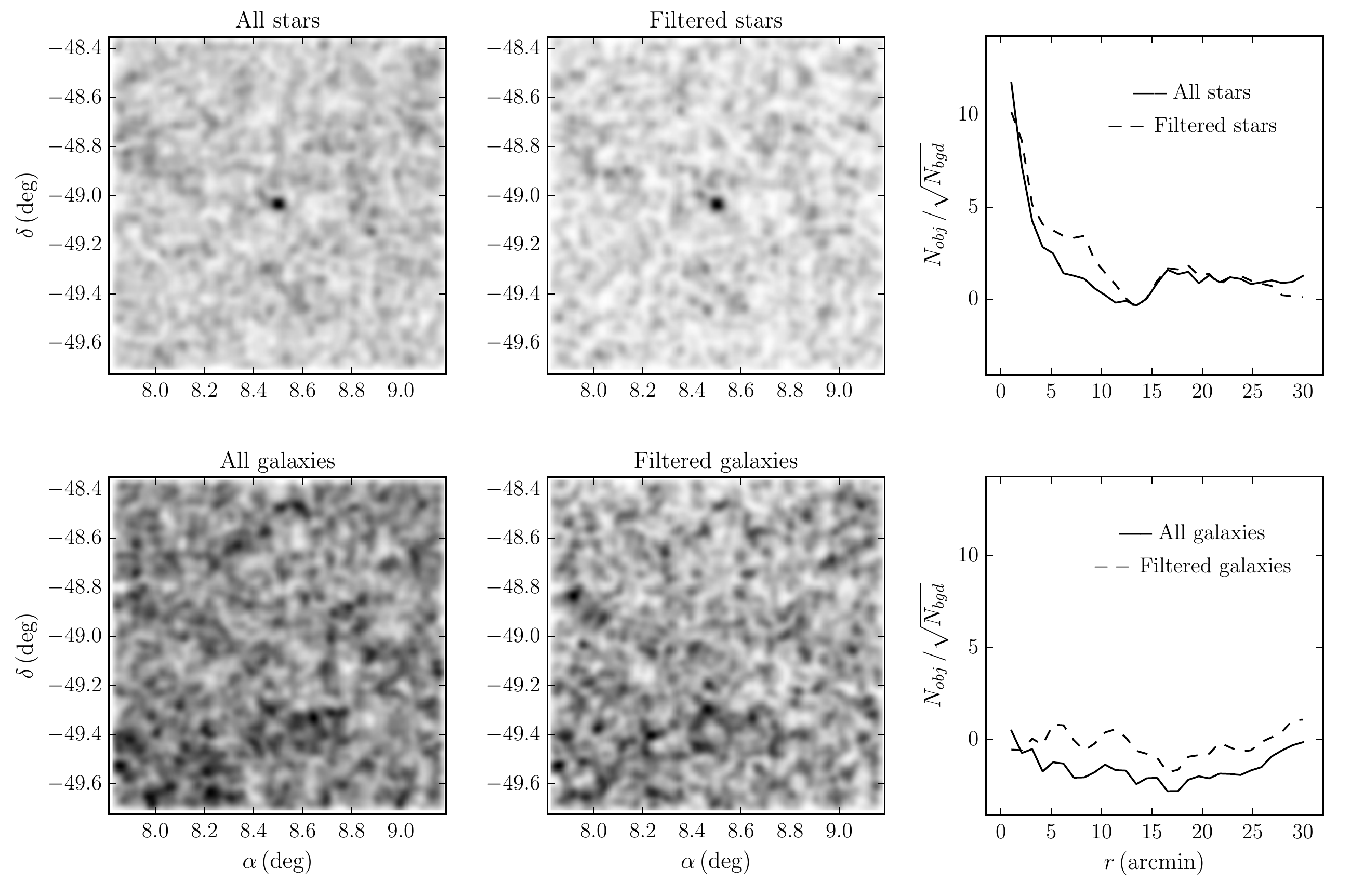}
\caption{\small Top left panel: on-sky number density map of stellar sources around candidate DES\,1. All stars are included. Top middle panel: similar to previous panel, but now only stars which lie close to the best fitting isochrone shown in Figure \ref{J0034cmd} are used. Top right panel: Significance as a function of radial distance from the centre of DES\,1. The solid (dashed) line correspond to all (isochrone filtered) stars as indicated. The corresponding panels at the bottom show the same plots but using the distribution of sources classified as galaxies.} 
\label{J0034map}
\end{center}
\end{figure*}

\section{DES\,1}
\label{sec:results}
DES\,1 stands out as the most conspicuous new candidate from our search. It is also directly seen as an overdensity of blue stellar sources in the DES coadd images (Figure \ref{J0034tv}). In Figure \ref{J0034map} we show the number density of stars on the sky around this object (top panels). The left panel shows all classified stellar sources, as described in \S \ref{sec:data}, and the middle one shows only those close to the best-fit isochrone (see below). A clear overdensity is seen in both. The Poisson significance profile shows a very pronounced peak at about $1\farcm 0$ from the object centre. It is built by taking the ratio of the number of stars internal to each radius $r$ and in excess of the background ($N_{bgd}$), $N_{obj}$, to the expected fluctuation in the same background. $N_{obj}=(N_{obs} - N_{bgd})$, where $N_{obs}$ is the total number of observed stars. $N_{bgd}$ is computed at an area ring at $30\farcm 0<r < 34\farcm 0 $ from the center of DES\,1. The bottom panels of  Figure \ref{J0034map} show that there is not an overdensity of galaxies at the position of DES\,1, and therefore it is unlikely that misclassified galaxies can account for the apparent stellar overdensity.

We use a maximum-likelihood technique to infer structural and CMD parameters for DES\,1. For simplicity, we assume that this object follows an exponential spatial profile with five free parameters: central RA ($\alpha_0$) and DEC ($\delta_0$), position angle $\theta$, ellipticity $\epsilon$, and exponential scale $r_e$. We follow the same method as \cite{Martin2008} to find the best-fit solution and the parameter uncertainties. We then use the CMD of the most likely members of the system to fit an isochrone model, whose free parameters are: age, $\mathrm{[Fe/H]}$, $(m-M)_0$, and $A_V$. The method is based on finding the peak likelihood in a series of model grids, as described in detail by Pieres et al (2015), in prep. 

\begin{figure*}
\begin{center}
\includegraphics[width=.71\textwidth]{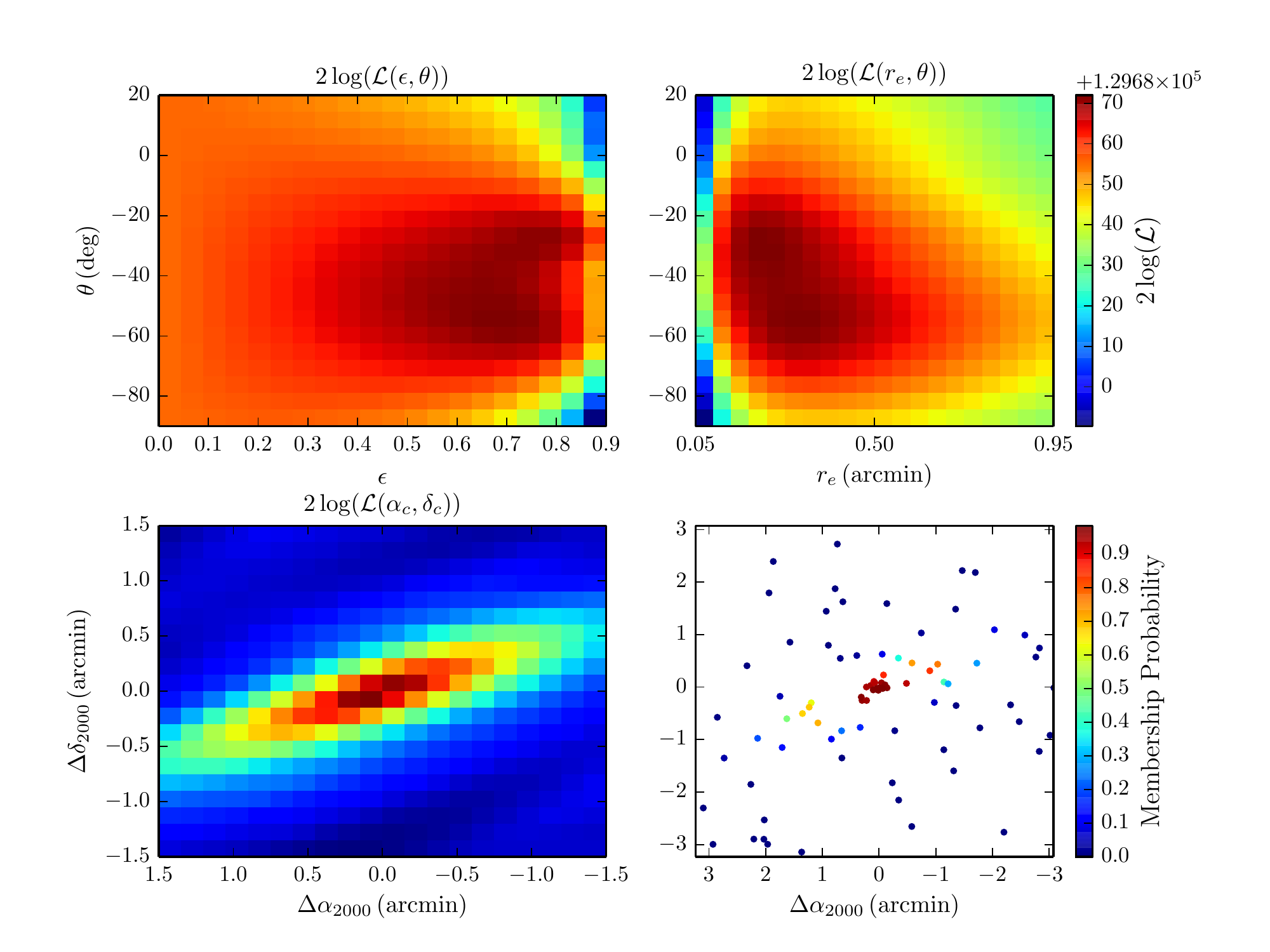}
\caption{Upper left panel: likelihood map for DES\,1 projected onto the position angle and eccentricity plane. Upper right: likelihood map for DES\,1 projected onto the position angle and exponential scale plane. Lower left: likelihood map for DES\,1 projected onto the central equatorial coordinates plane. Lower right: Spatial map of stars with probability larger than 1\% to belong to DES\,1 color-coded by probability. The best-fit parameters are listed in Table \ref{fitpars}.
}
\label{J0034prof}
\end{center}
\end{figure*}

\begin{figure}
\centering
\includegraphics[width=.45\textwidth]{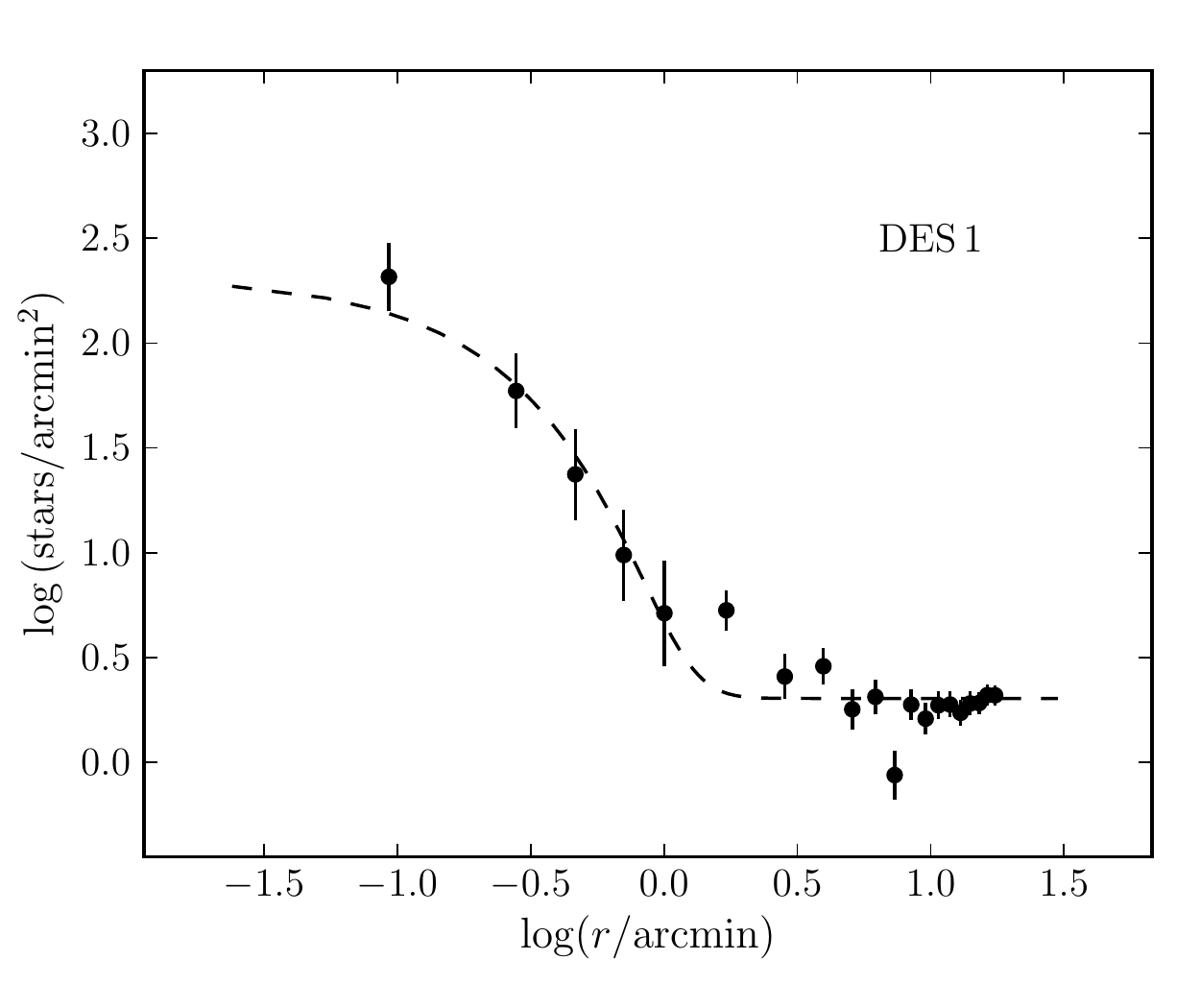}
\caption{\small Solid points show a binned version of the radial density profile of DES\,1. The error bars are 1-sigma Poisson. The dashed line is the best maximum-likelihood fit model to an exponential profile.}
\label{J0034profbin}
\end{figure}

In Figure \ref{J0034prof}, we show the results of the profile fit to DES\,1. The first three panels show the likelihood values projected on individual planes of this five-dimensional space, which all show well-defined peaks. The corresponding parameter values and their uncertainties (computed as discussed in \cite{Martin2008} and Pieres et al 2015, in prep.) are listed in Table \ref{fitpars}. The last panel shows the individual stars coded by their membership probabilities. We note that DES\,1 is a quite elongated object ($\epsilon \simeq 0.7$). Figure \ref{J0034profbin} shows a binned radial density profile compared to the best fit exponential model. The central density of DES\,1 is $\mathrm{\simeq 200\,stars/arcmin^2}$. A clear excess of stars relative to the background is seen out to $\mathrm{\simeq 2\farcm 0}$. In fact, a systematic excess of stars relative the best-fit model is seen beyond $\mathrm{\simeq 1\farcm 0}$, which may indicate that a King profile may better describe the observed profile.

\begin{figure*}
\begin{center}
\includegraphics[width=.95\textwidth]{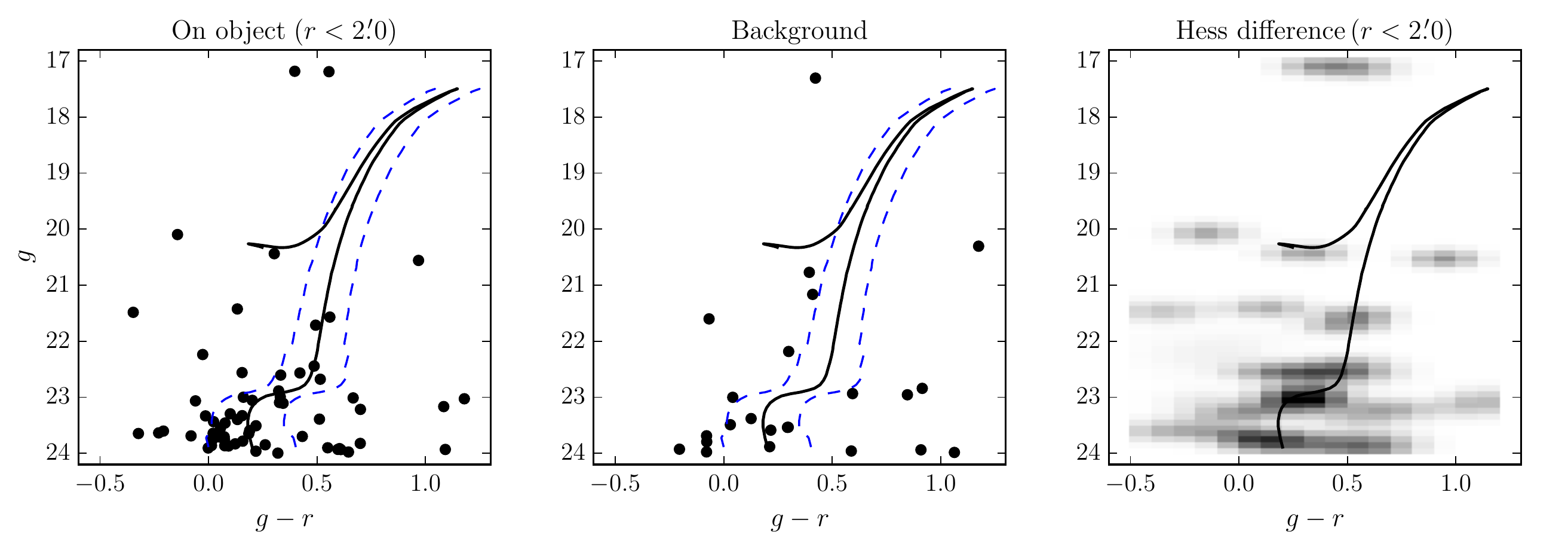}
\caption{\small Left panel: CMD of stars within $2\farcm 0$ from the centre of
DES\,1. The maximum-likelihood isochrone fit is shown, along with ridge lines meant to bracket the most likely members. Middle panel: CMD of background stars in an equal area on the sky as the previous panel. Right panel: Hess diagram of the CMD difference between the two previous panels.}   
\label{J0034cmd}
\end{center}
\end{figure*}

The CMD for this candidate is shown on the left panel of Figure \ref{J0034cmd}. Only stars inside the radius corresponding to the peak in the Poisson significance profile shown in Figure \ref{J0034map} are shown. The middle panel shows the field CMD in a ring of equal area, whose inner radius equals ten times the radius of the object's CMD. The best-fit isochrone is also shown; parameter values are listed in Table \ref{fitpars}. Also shown are the sequences bracketing the best-fit isochrone fit at a distance of $\sqrt{0.1^2 + {\tt MAG\_ERR}^{2} + {\tt COL\_ERR}^{2}}$ on the CMD plane, where ${\tt MAG\_ERR}$ and ${\tt COL\_ERR}$ are the mean photometric uncertainties along the CMD axes. We use the CMD space between them to filter the most likely cluster stars. See the middle and right panels of Figure \ref{J0034map} for a density map and a Poisson significance profile of objects inside this isochrone filter, respectively,. The CMD difference relative to the background field is shown as the Hess diagram in the right panel of Figure  \ref{J0034cmd}. The main sequence turn-off (MSTO) and sub-giant branch (SGB) are clearly visible.

As previously mentioned, we summarize the inferred properties of DES\,1 in Table \ref{fitpars}. The table lists positions, structural parameters, central ($\sigma_c$) and background ($\sigma_{bg}$) densities, half-light radius ($r_h$), distance ($D_\sun$), Absolute Magnitude ($M_V$), Test Statistic ($\mathrm{TS}$), and peak Poisson significance ($\mathrm{PS}$) value, as well as best-fit CMD parameters. The $\mathrm{TS}$ is based on the likelihood ratio between a hypothesis that includes an object versus a field-only hypothesis (see \citealt{Bechtol2015}).
 
A maximum-likelihood approach that simultaneously fits the profile (assuming a Plummer model) and the distance (but assuming an age of $\mathrm{12\,Gyrs}$ and a spread in metallicities) has also been tried. It yields a distance modulus of $(m-M)_0 = 19.6$, in agreement with the method described earlier. The alternative $r_h$, however, is larger, $r_h\simeq\mathrm{1\farcm 0}$. 
Visual fits were also independently made to the object's CMD. Again, the results agree well with those from the maximum-likelihood fits shown on the table: $\mathrm{\log (Age) = 9.9}$, $\mathrm{[Fe/H] = -1.98}$, $A_V=0.03$, and $(m-M)_0 = 19.9$. 
The quoted $M_V$ estimate was computed by integrating over all masses along the best fit model isochrone assuming a \cite{Kroupa2001} IMF, and normalizing the number of objects by those observed in the CMD with $r < 23\,\mathrm{mag}$ and which fall in the isochrone filter. We convert from DES $g$ and $r$ magnitudes  to $V$ magnitudes using
\begin{align}
g_{\rm SDSS}&= g_{\rm DES} + 0.104(g_{\rm DES} - r_{\rm DES}) -0.01\nonumber\\
r_{\rm SDSS}&= r_{\rm DES} + 0.102(g_{\rm DES} - r_{\rm DES}) -0.02\\
V&= g_{\rm SDSS}-0.59(g_{\rm SDSS}-r_{\rm SDSS})-0.01.\nonumber\label{eq:8}
\end{align}
These transformations were derived by \cite{Jester2005} using a SDSS stellar calibration sample.

\begin{figure*}
\begin{center}
\includegraphics[width=.94\textwidth]{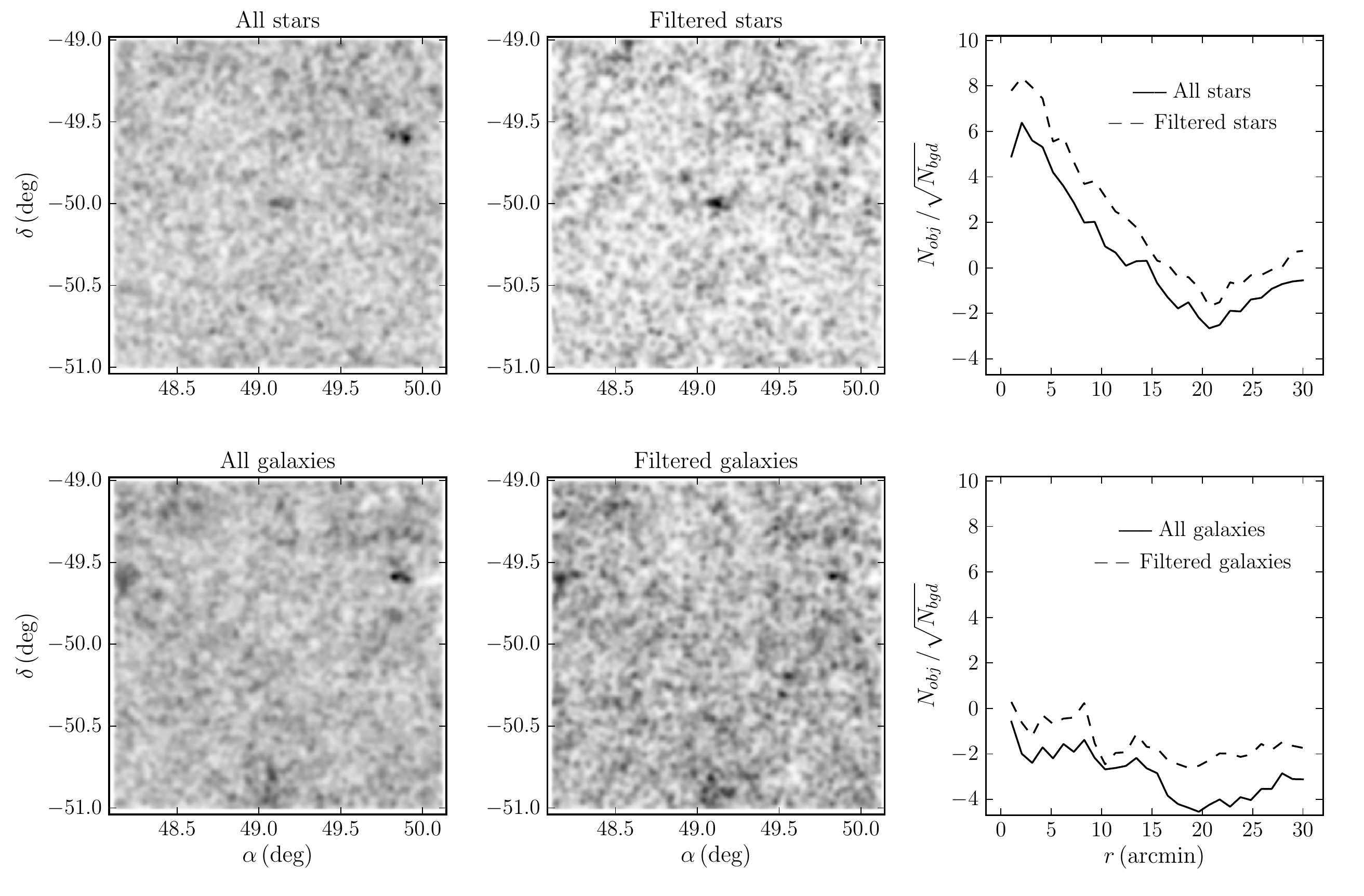}
\caption{\small All panels are the same as those in Fig. \ref{J0034map} but now for the Horologium\,II satellite dwarf.} 
\label{HorIImap}
\end{center}
\end{figure*}

\begin{table}
\caption{Properties of DES\,1}\centering
\scalebox{1.0}{
\begin{tabular}{lcc}\hline
Parameters && DES\,1\\\hline
$\alpha_0\,(J2000)$&& $\mathrm{00^h33^m59.7^s\pm 1.7^s}$\\
$\delta_0\,(J2000)$&& $\mathrm{-49\degr 02\arcmin 20.0\arcsec \pm 13.7\arcsec}$\\
$r_e$&& $0\farcm 23 \pm 0\farcm 10$\\
$\theta\,\mathrm{(deg)}$&& $-57.9 \pm 22.2$\\
$\epsilon$&& $0.69\pm 0.22$\\
$\sigma_c\,(\mathrm{stars/arcmin^2})$&& $204.98 \pm 35.81$\\
$\sigma_{\mathrm{bg}}\,(\mathrm{stars/arcmin^2})$&& $2.02 \pm 0.01$\\
$r_h$&& $0\farcm 39 \pm 0\farcm 17$\\
$D_\sun\,(\mathrm{kpc})$&&$\sim 87.1$\\
$r_h\,\mathrm{(pc)}$&& $9.88 \pm 4.31$\\
$M_V$&&$-3.05_{-0.42}^{+0.69}$\\
$\mathrm{TS}$&& $134.7$\\
$\mathrm{PS}$&& $11.7 \pm 3.1$\\\hline
$\mathrm{[Fe/H]}$&& $-1.98$\\
$\log(\mathrm{Age})$&& $10.00$\\
$A_V$&& $0.0$\\
$(m-M)_0$&& $19.70$\\\hline
\end{tabular}}\label{fitpars}
\end{table}

\section{Horologium\,\,II}
\label{sec:HorII}
As mentioned earlier, \cite{Kim2015b} report on the identification of an extra MW satellite galaxy candidate besides those found by \cite{Bechtol2015} and \cite{Koposov2015}. Our reanalysis of Y1A1 presented here confirms this object, Horologium\,II, as a real stellar system. In fact, once we allow for a deeper magnitude threshold, as explained in \S \ref{sec:data}, we detect it not only with the method described in \S \ref{sec:method} but also with the maximum-likelihood satellite search method detailed out in \cite{Bechtol2015}, 

Figure \ref{HorIImap} shows the same information as Figure \ref{J0034map}, but now for Horologium\,II. A clear overdensity of stars is seen in the density map on the sky. The statistical significance of this overdensity is close to 8 times the expected Poisson fluctuation in the background.

Using the same maximum-likelihood fit method to the on-sky and CMD distributions as in the previous section, we obtain the following structural and isochrone parameters: $(m-M)_0 \simeq 19.5$ (heliocentric distance $D \simeq 79.4$ kpc), $\mathrm{\log (Age) \simeq 9.9}$, $\mathrm{[Fe/H] \simeq -1.2}$, $A_V \simeq 0.01$, $r_h \simeq 2.2 \arcmin$ ($51.2$ pc), position angle $\theta \simeq 106 \deg$, elipticity $\epsilon \simeq 0.6$, and absolute magnitude $M_V \simeq -3.1$. 

Our distance, size and flatenning estimates are in very good agreement (within $1 \sigma$) with those from the discovery paper by \cite{Kim2015b}. The position angle and absolute luminosity are within $2 \sigma$ of their quoted values. The largest discrepancies occur for the isocrhone parameters. We fit the observed CMD of Horologium\,II to a younger and more metal rich PARSEC model than \cite{Kim2015b}. However, our likelihood function over the metallicity and age plane exhibits a tail of high likelihood values towards lower $Z$ and older ages. The discrepancy may also be related to the fact that our listed redenning value comes out of the maximum-likelihood CMD fit, whereas the values from \cite{Schlegel1998} maps \{with corrections from \cite{Schlafly2011}\} are used in the discovery paper.

\section{Conclusions}
\label{sec:conclusions}

In this paper we  make a deeper probe on the DES Y1A1 catalog in search for additional Galactic satellites besides those previously reported by the collaboration \citep{Bechtol2015}. We report the discovery of a new stellar system in the MW halo, using catalogs based on first-year data from the Dark Energy Survey. We have explored the data at least $\mathrm{1\,mag}$ deeper ($g<24\,\mathrm{mag}$) than previously done in \cite{Bechtol2015}. The candidate adds to the ten previously identified systems found using DECam images by \cite{Bechtol2015,Koposov2015,Kim22015,Kim2015b}.  We also confirm the dwarf galaxy candidate Horologium\,II, originally discovered by \cite{Kim2015b}, as a significant overdensity in the Y1A1 catalog. Our best fit structural parameters to this later are in general agreement with the ones derived by those authors, although the isochrone fit tends to point to a younger and more metal-rich object than previously reported.

DES\,1 is detected as a significant stellar overdensity both spatially and on the CMD plane. Isochrone fits based on two very different methods show that it is made up of old and metal poor stars, as commonly observed in MW satellites found in the Galactic halo. Extinction towards the object, as estimated from the \cite{Schlegel1998} maps, is found to be relatively small ($\mathrm{\sim 0.03\,mag}$). 

Maximum-likelihood fits of the spatial profile of DES\,1 yield a scale radius of $r_e \simeq 0\farcm 2$, which at a distance $\simeq 87\,\mathrm{kpc}$ corresponds to a physical size of $r_h\simeq 9.88\,\mathrm{pc}$. Its estimated distance places this faint cluster candidate as one of farthest away from the Sun. The absolute magnitude has been determined using a similar approach as \cite{Koposov2015}, yielding $M_V \simeq -3.05$ for DES\,1. Taken together, the physical size and luminosity place DES\,1 in the locus occupied by low-luminosity star clusters. 
DES\,1 is also significantly elongated ($\epsilon \simeq 0.7$), something that is apparent not only from the profile fit, but also from the distribution of the stars on the sky (Figures \ref{J0034tv} and \ref{J0034prof}). 
It is, in fact, the most elongated halo cluster known to date, although, given the errorbars, its excentricity is marginally consistent with those of Kim\,1, Laevens\,3 (see \citealt{Kim2015a, Laevens2015b}, respectively). The very high inferred excentricity suggests that DES\,1 is in dynamical process of tidal disruption, despite its large distance, and makes it a very interesting object for deeper imaging and spectroscopic follow up.

Due to the low number of probable member stars detected in the DES imaging, it is difficult to extract more reliable information about DES\,1 at this stage. Steps to acquire deeper imaging of this object are already under way.\par
A search for satellites in data collected by the Dark Energy Survey during its second season, including new areas sky, is under way. It is likely that additional new stellar systems will be discovered soon.

\section*{Acknowledgements} 
This paper has gone through internal review by the DES
collaboration.\par
 Funding for the DES Projects has been provided by the U.S. Department of Energy, the U.S. National Science Foundation, the Ministry of Science and Education of Spain, 
the Science and Technology Facilities Council of the United Kingdom, the Higher Education Funding Council for England, the National Center for Supercomputing 
Applications at the University of Illinois at Urbana-Champaign, the Kavli Institute of Cosmological Physics at the University of Chicago, 
the Center for Cosmology and Astro-Particle Physics at the Ohio State University,
the Mitchell Institute for Fundamental Physics and Astronomy at Texas A\&M University, Financiadora de Estudos e Projetos, 
Funda{\c c}{\~a}o Carlos Chagas Filho de Amparo \`a Pesquisa do Estado do Rio de Janeiro, Conselho Nacional de Desenvolvimento Cient{\'i}fico e Tecnol{\'o}gico and the Minist{\'e}rio da Ci{\^e}ncia, Tecnologia e Inova{\c c}{\~a}o, the Deutsche Forschungsgemeinschaft and the Collaborating Institutions in the Dark Energy Survey. 
The DES data management system is supported by the National Science Foundation under Grant Number AST-1138766.
The DES participants from Spanish institutions are partially supported by MINECO under grants AYA2012-39559, ESP2013-48274, FPA2013-47986, and Centro de Excelencia Severo Ochoa SEV-2012-0234, 
some of which include ERDF funds from the European Union.

The Collaborating Institutions are Argonne National Laboratory, the University of California at Santa Cruz, the University of Cambridge, Centro de Investigaciones En{\'e}rgeticas, 
Medioambientales y Tecnol{\'o}gicas-Madrid, the University of Chicago, University College London, the DES-Brazil Consortium, the University of Edinburgh, 
the Eidgen{\"o}ssische Technische Hochschule (ETH) Z{\"u}rich, 
Fermi National Accelerator Laboratory, the University of Illinois at Urbana-Champaign, the Institut de Ci\`encies de l'Espai (IEEC/CSIC), 
the Institut de F{\'i}sica d'Altes Energies, Lawrence Berkeley National Laboratory, the Ludwig-Maximilians Universit{\"a}t M{\"u}nchen and the associated Excellence Cluster Universe, 
the University of Michigan, the National Optical Astronomy Observatory, the University of Nottingham, The Ohio State University, the University of Pennsylvania, the University of Portsmouth, 
SLAC National Accelerator Laboratory, Stanford University, the University of Sussex, and Texas A\&M University.

The DES data management system is supported by the National Science Foundation under Grant Number AST-1138766.
The DES participants from Spanish institutions are partially supported by MINECO under grants AYA2012-39559, ESP2013-48274, FPA2013-47986, and Centro de Excelencia Severo Ochoa SEV-2012-0234.\par
Research leading to these results has received funding from the European Research Council under the European Union’s Seventh Framework Programme (FP7/2007-2013) including ERC grant agreements 
 240672, 291329, and 306478.



\bibliographystyle{mn2e}          
\bibliography{bib.bib}\vspace*{.4cm}
{\small\it
$^1$Instituto de F\'\i sica, UFRGS, Caixa Postal 15051, Porto Alegre, RS - 91501-970, Brazil\\
$^2$Laborat\'orio Interinstitucional de e-Astronomia - LIneA, Rua Gal. Jos\'e Cristino 77, Rio de Janeiro, RJ - 20921-400, Brazil\\
$^3$Observat\'orio Nacional, Rua Gal. Jos\'e Cristino 77, Rio de Janeiro, RJ 20921-400, Brazil \\
$^4$Kavli Institute for Cosmological Physics, University of Chicago, Chicago, IL 60637, USA\\
$^5$Fermi National Accelerator Laboratory, P. O. Box 500, Batavia, IL 60510, USA\\
$^6$Department of Physics, University of Surrey, Guildford GU2 7XH, UK\\
$^{7}$ Cerro Tololo Inter-American Observatory, National Optical Astronomy Observatory, Casilla 603, La Serena, Chile\\
$^{8}$Department of Physics \& Astronomy, University College London, Gower Street, London, WC1E 6BT, UK\\
$^{9}$CNRS, UMR 7095, Institut d'Astrophysique de Paris, F-75014, Paris, France\\
$^{10}$Sorbonne Universit\'es, UPMC Univ Paris 06, UMR 7095, Institut d'Astrophysique de Paris, F-75014, Paris, France\\
$^{11}$Kavli Institute for Particle Astrophysics \& Cosmology, P. O. Box 2450, Stanford University, Stanford, CA 94305, USA\\
$^{12}$SLAC National Accelerator Laboratory, Menlo Park, CA 94025, USA\\
$^{13}$Department of Astronomy, University of Illinois, 1002 W. Green Street, Urbana, IL 61801, USA\\
$^{14}$National Center for Supercomputing Applications, 1205 West Clark St., Urbana, IL 61801, USA\\
$^{15}$Institut de Ci\'encies de l'Espai, IEEC-CSIC, Campus UAB, Carrer de Can Magrans, s/n,  08193 Bellaterra, Barcelona, Spain\\
$^{16}$Institut de F\'{\i}sica d'Altes Energies, Universitat Aut\'onoma de Barcelona, E-08193 Bellaterra, Barcelona, Spain\\
$^{17}$Excellence Cluster Universe, Boltzmannstr.\ 2, 85748 Garching, Germany\\
$^{18}$Faculty of Physics, Ludwig-Maximilians University, Scheinerstr. 1, 81679 Munich, Germany\\
$^{19}$Department of Physics and Astronomy, University of Pennsylvania, Philadelphia, PA 19104, USA\\
$^{20}$Jet Propulsion Laboratory, California Institute of Technology, 4800 Oak Grove Dr., Pasadena, CA 91109, USA\\
$^{21}$Kavli Institute for Cosmological Physics, University of Chicago, Chicago, IL 60637, USA\\
$^{22}$Department of Physics, University of Michigan, Ann Arbor, MI 48109, USA.\\
$^{23}$Max Planck Institute for Extraterrestrial Physics, Giessenbachstrasse, 85748 Garching, Germany\\
$^{24}$University Observatory Munich, Scheinerstrasse 1, 81679
Munich, Germany\\
$^{25}$Center for Cosmology and Astro-Particle Physics, The Ohio State University, Columbus, OH 43210, USA\\
$^{26}$Department of Physics, The Ohio State University, Columbus, OH 43210, USA\\
$^{27}$Australian Astronomical Observatory, North Ryde, NSW 2113, Australia\\
$^{28}$George P. and Cynthia Woods Mitchell Institute for Fundamental Physics and Astronomy, and Department of Physics and Astronomy, Texas A\&M University, College Station, TX 77843, USA\\
$^{29}$Department of Astronomy, The Ohio State University, Columbus, OH 43210, USA\\
$^{30}$Instituci\'o Catalana de Recerca i Estudis Avan\c{c}ats, E-08010 Barcelona, Spain.\\
$^{31}$Institute of Cosmology \& Gravitation, University of Portsmouth, Portsmouth, PO1 3FX, UK.\\
$^{32}$Department of Physics and Astronomy, Pevensey Building, University of Sussex, Brighton, BN1 9QH, UK.\\
$^{33}$Centro de Investigaciones Energ\'eticas, Medioambientales y Tecnol\'ogicas (CIEMAT), Madrid, Spain\\
$^{34}$Department of Physics, University of Illinois, 1110 W. Green St., Urbana, IL 61801, USA
}
\label{lastpage} 
\end{document}